\documentstyle[twoside,fleqn,espcrc2]{article}
\include{psfig}

\newcommand{\AmS}{{\protect\the\textfont2
  A\kern-.1667em\lower.5ex\hbox{M}\kern-.125emS}}
\newcommand{\vp}{\vec{p}}
\newcommand{\vs}{\vec{s}}
\newcommand{\vx}{\vec{x}}
\newcommand{\ve}{\vec{e}}
\newcommand{\vz}{\vec{0}}

\title{The Spin Content of the Nucleon%
       \thanks{Talk presented by G. Schierholz}}
\author{R. Altmeyer%
        \address{Deutsches Elektronen-Synchrotron DESY, \\
         Notkestra{\ss}e 85, W-2000 Hamburg 52, Germany},
        M. G\"ockeler%
        \address{Institut f{\"u}r Theoretische Physik, RWTH Aachen,\\
         Sommerfeldstra{\ss}e, W-5100 Aachen, Germany}%
        \address{Gruppe Theorie der Elementarteilchen,  \\
         H{\"o}chstleistungsrechenzentrum HLRZ,  \\
         c/o Forschungszentrum J{\"u}lich, W-5170 J{\"u}lich,
                                                             Germany},
        R. Horsley$^{\rm c}$,
        E. Laermann%
        \address{Fakult{\"a}t f{\"u}r Physik, Universit{\"a}t
         Bielefeld, \\
         Postfach 8640, W-4800 Bielefeld, Germany}
        and
        G. Schierholz$^{{\rm a,c}}$}

\begin{document}

\begin{abstract}
The fraction of the nucleon spin that is
carried by the quarks, $\Delta \Sigma$,
is computed in lattice QCD with dynamical staggered fermions. We obtain
the value $\Delta \Sigma = 0.18 \pm 0.02$.
\end{abstract}

\maketitle

% reset footnote counter
\setcounter{footnote}{0}

\section{INTRODUCTION}

The EMC measurement \cite{EMC} of the spin-dependent structure function
of the proton, $g_1 (x, Q^2)$, has provoked many
speculations \cite{many} about the internal spin structure of the
nucleon. The first moment of $g_1 (x, Q^2)$, which through the
operator product expansion is given by the proton matrix element
of the axial vector current weighted by the square of the quark
charges (modulo radiative corrections), was found to be
\begin{eqnarray}
\begin{array}{ll}
\int_0^1 dx g_1 (x, Q^2) & = \frac{1}{2} (\frac{4}{9} \Delta u
+ \frac{1}{9} \Delta d + \frac{1}{9} \Delta s) \\[0.5em]
                         & = 0.126 \pm 0.010 \pm 0.015
\end{array}
\end{eqnarray}
with
\begin{equation}
\Delta q \: s_\mu =
<{\vp},s| \bar{q} \gamma_\mu \gamma_5 q|{\vp},s>,
\, q = u, d, s,
\end{equation}
where $s_\mu$ is the covariant spin vector of the proton. The average
value of $Q^2$ of the EMC data is $\approx 11$ GeV$^2$.
If we combine the result (1)
with information from hyperon decays, neutron
$\beta$-decay and the assumption of flavor $SU(3)$, we
obtain \cite{Ellis}
\begin{equation}
\Delta \Sigma = \Delta u + \Delta d + \Delta s = 0.04 \pm 0.16.
\end{equation}
The quantity $\Delta \Sigma$ is the axial baryonic charge of the
nucleon,
\begin{eqnarray}
\begin{array}{l}
\Delta \Sigma \: s_\mu  =  \\[0.5em]
<{\vp}, s|[ \bar{u} \gamma_\mu \gamma_5 u +
\bar{d} \gamma_\mu \gamma_5 d + \bar{s} \gamma_\mu \gamma_5 s ]
|{\vp}, s>  \\[0.5em]
 = <{\vp}, s| j_\mu^{0 5} |{\vp}, s>,
\end{array}
\end{eqnarray}
which in a naive wave function picture can be interpreted as the
fraction of the nucleon spin that is carried by the quarks. For
example, in a $SU(6)$-type model of the nucleon we would obtain
$\Delta \Sigma = 1$. The vanishingly small experimental value of
$\Delta \Sigma$ is referred to as the spin crisis of the nucleon.

The nucleon matrix element of the axial baryonic current can be
written \cite{Fritzsch}
\begin{eqnarray}
\begin{array}{l}
<{\vp},s| j_\mu^{0 5}|{\vp}^{\, \prime}, s^\prime> = \\[0.5em]
\bar{u}({\vp}, s) [G_1(k^2)\gamma_\mu\gamma_5 - G_2(k^2) k_\mu%
\gamma_5] u({\vp}^{\, \prime}, s^\prime),
\end{array}
\end{eqnarray}
where $k$ is the momentum transfer and
\begin{equation}
G_1(0) = \Delta \Sigma.
\end{equation}
Unlike the octet current, $j_\mu^{0 5}$ is not conserved due to the
anomaly. In the chiral limit
\begin{equation}
\partial_\mu j_\mu^{0 5} = N_f \frac{1}{8 \pi^2} {\rm Tr}
F_{\mu \nu} \tilde{F}_{\mu \nu}, \; \tilde{F}_{\mu \nu} =
\frac{1}{2} \epsilon_{\mu \nu \rho \sigma} F_{\rho \sigma},
\end{equation}
where $N_f$ is the number of flavors. As a result, the form factor
$G_2(k^2)$ does not have a (Goldstone boson) pole at $k^2 = 0$.
Writing
\begin{eqnarray}
\begin{array}{l} \displaystyle
<{\vp}, s| N_f \frac{1}{8 \pi^2} {\rm Tr} F_{\mu \nu}
\tilde{F}_{\mu \nu} |{\vp}^{\, \prime}, s^\prime> \\[0.5em]
= 2 m_N A(k^2) \: \bar{u}({\vp}, s)
i \gamma_5 u({\vp}^{\, \prime}, s^\prime),
\end{array}
\end{eqnarray}
where $m_N$ is the nucleon mass, one then finds
\begin{equation}
A(0) = G_1(0) \equiv \Delta \Sigma.
\end{equation}
Thus, the quark spin fraction of the nucleon spin is given by the matrix
element of the anomalous divergence of the axial baryonic current.

\section{LATTICE CALCULATION}

A meaningful lattice calculation of $\Delta \Sigma$ requires (i) the
inclusion of dynamical fermions and (ii) a proper definition of
${\rm Tr} F_{\mu \nu} \tilde{F}_{\mu \nu}$ in order to account for the
topological origin of the anomalous divergence. None of these
requirements were fulfilled in earlier calculations \cite{Mandula}.

The calculations in this paper are done on a $16^3 \cdot 24$ lattice
at $\beta = 5.35$ and $m a = 0.01$
\footnote{In some cases we explicitly
state the lattice spacing $a$ in order to avoid confusion.} with four
flavors of dynamical staggered fermions using the hybrid Monte Carlo
algorithm. Our data sample consists of 85 configurations separated by
five trajectories. These configurations were used in ref. \cite{MTC}
to compute the hadron mass spectrum. The lattice parameters correspond
to a renormalization group invariant quark mass of $m^{RGI} = 35$ MeV
in the $\overline{MS}$ scheme at a scale of 1 GeV.

For ${\rm Tr} F_{\mu \nu} \tilde{F}_{\mu \nu}$ we use L\"uscher's
definition \cite{Luscher},
\begin{eqnarray}
\begin{array}{l}
{\rm Tr} F_{\mu \nu} \tilde{F}_{\mu \nu}(x) = \frac{2}{3} \sum_{\mu \nu
\rho \sigma} \epsilon_{\mu \nu \rho \sigma} \\[0.5em]
\times \; \{ 3 \int_{p(x+\hat{\mu}+\hat{\nu},\mu,\nu)} d^2 z \\[0.5em]
\times \;{\rm Tr} \:
[P^x_{x+\hat{\mu}+\hat{\nu},\mu \nu} \partial_\rho P^{x \; -1}_{x+
\hat{\mu}+\hat{\nu},\mu \nu} \\[0.5em]
\times \; R^{x \; -1}_{x+\hat{\mu},\mu;\nu}
\partial_\sigma R^x_{x+\hat{\mu},\mu;\nu}] \\[0.5em]
- \; 3 \int_{p(x+\hat{\nu},\mu,\nu)} d^2 z \: {\rm Tr} \:
[P^x_{x+\hat{\nu},\mu \nu} \partial_\rho P^{x \; -1}_{x+
\hat{\nu},\mu \nu} \\[0.5em]
\times \; R^{x \; -1}_{x,\mu;\nu}
\partial_\sigma R^x_{x,\mu;\nu}] \\[0.5em]
- \; \int_{f(x+\hat{\mu},\mu)} d^3 z \: {\rm Tr} \:
[S^x_{x+\hat{\mu},\mu} \partial_\nu S^{x \; -1}_{x+
\hat{\mu},\mu} \\[0.5em]
\times \; S^{x}_{x+\hat{\mu},\mu}
\partial_\rho S^{x \; -1}_{x+\hat{\mu},\mu} \;
       S^{x}_{x+\hat{\mu},\mu}
\partial_\sigma S^{x \; -1}_{x+\hat{\mu},\mu}] \\[0.5em]
+ \; \int_{f(x,\mu)} d^3 z \: {\rm Tr} \:
[S^x_{x,\mu} \partial_\nu S^{x \; -1}_{x
,\mu} \\[0.5em]
\times \; S^{x}_{x,\mu}
\partial_\rho S^{x \; -1}_{x,\mu} \;
       S^{x}_{x,\mu}
\partial_\sigma S^{x \; -1}_{x,\mu}]\},
\end{array}
\end{eqnarray}
where $P$, $R$ and $S$ are certain parallel transporters extrapolated to
the interior of the plaquettes, $z \in p$, and faces, $z \in f$.
This expression proceeds from the principle bundle which is
reconstructed from the lattice gauge field.
The resulting topological
charge,
\begin{equation}
Q = -\frac{1}{16 \pi^2} \sum_x {\rm Tr} F_{\mu \nu} \tilde{F}_{\mu \nu},
\end{equation}
assumes integer values as in the continuum. The major drawback of
eq. (10) is that it involves a
three-dimensional integral over the faces of the
hypercubes. The more elegant algorithm of Phillips and Stone \cite{P&S}
does not lead naturally to a charge density. It should be noted that
in the presence of light dynamical fermions the topological
susceptibility, $\chi_t = <Q^2>/V$ ($V$: volume), is not affected by
dislocations \cite{Teper,GKLSW}. In the large volume limit the minimal
action for a $|Q| = 1$ configuration becomes \cite{full}
\begin{equation}
S_{min} \propto \frac{V}{\beta} {\rm ln} (m a)^{-1},
\end{equation}
which grows beyond any bound for fixed bare mass
$m$ (in physical units).

According to eqs. (8) and (9), $\Delta \Sigma$ is given by
\begin{eqnarray}
\begin{array}{l} \displaystyle
\Delta \Sigma =
\lim_{{\vp} \rightarrow 0} \, i \frac{|{\vs}|}{{\vp \vs}} \\[0.7em]
\displaystyle
 \times <{\vp}, s| \frac{1}{2 \pi^2} {\rm Tr} F_{\mu \nu}
\tilde{F}_{\mu \nu} |{\vz}, s>.
\end{array}
\end{eqnarray}
On a finite lattice we cannot reach ${\vp} = 0$. We therefore shall
evaluate $\Delta \Sigma$ at the smallest (non-zero)
momentum transfer, i.e.
$|{\vp}| = 2 \pi/16$. This is of the order of the pion mass.
Choosing ${\vs}/|{\vs}| = \pm \, {\vp}/
|{\vp}|$, we thus have to compute
\begin{eqnarray}
\begin{array}{l} \displaystyle
C(t) = \pm \: \frac{i}{|{\vp}|} \; \times \\[0.7em]
\displaystyle
<B_{{\vp}}(t) \: P_\pm \: \frac{1}{2 \pi^2}%
{\rm Tr} F_{\mu \nu} \tilde{F}_{\mu \nu} (x_4) \: \bar{B}_{{\vz}}(0)>,
\end{array}
\end{eqnarray}
where $\bar{B}$, $B$ are the baryon creation and annihilation
operators, $P_\pm$ is the spin projection operator and
\begin{equation}
{\rm Tr} F_{\mu \nu} \tilde{F}_{\mu \nu}(x_4) =
\sum_{\vx} {\rm e}^{i {\vp \vx}} \: {\rm Tr} F_{\mu \nu}
\tilde{F}_{\mu \nu}(x).
\end{equation}
Equation (14) contains 12 independent correlation functions.
Averaging over all of them gives \footnote{Note that the matrix elements
between $N$ and $\Lambda$ vanish in this case.}
\begin{eqnarray}
\begin{array}{l} \displaystyle
C(t) =
\Delta \Sigma \: A_+ \, {\rm e}^{-m_N x_4 -E_N (t-x_4)} \\[0.5em]
\displaystyle
+ \Delta \Sigma_\Lambda \: A_{-} \, (-1)^t {\rm e}^{-m_\Lambda x_4%
-E_\Lambda (t-x_4)} + \cdots ,
\end{array}
\end{eqnarray}
where $\Delta \Sigma_\Lambda$ ($m_\Lambda$)
is the quark spin fraction (mass) of the
$\Lambda$, the opposite parity partner of the nucleon \cite{MTC}, and
$E_{N,\Lambda} = \sqrt{{\vp}{\, ^2} + m_{N,\Lambda}^2}$.
The amplitudes $A_{+}$,
$A_{-}$ are those of the ordinary
correlation function $<B(t) \, \bar{B}(0)>$.
The dots in
eq. (16) stand for contributions from higher excitations and from
baryons backtracking around the boundary.

A non-trivial problem is the computation of ${\rm Tr} F_{\mu \nu}
\tilde{F}_{\mu \nu}$ for a given gauge field configuration.
For gauge group $SU(2)$ we could do one integral
in ${\rm Tr} F_{\mu \nu} \tilde{F}_{\mu \nu}$ analytically \cite{Fox},
which made its computation just feasible. In the present case of gauge
group $SU(3)$ we shall make use of the fact that the calculation can be
reduced to the case of $SU(2)$ by means of the so-called reduction of
the structure group \cite{Choquet,GLSW}. The argument goes as follows.
Since ${\rm Tr} F_{\mu \nu} \tilde{F}_{\mu \nu} = \partial_\mu
K_\mu$, we can write $s_0 \, G_1(0) \propto <{\vp}, s| K_0 |{\vp}, s>$.
It turns out \cite{GLSW} that $\sum_{{\vx}} K_0 =
\sum_{{\vx}} \partial_i K_{0 i} \in {\bf Z}$ is a topological
invariant. As in the case of the topological charge we then may
continuously deform the $SU(3)$ gauge field to a $SU(2)$ gauge field
without changing $\sum_{{\vx}} K_0$. The reason is that
$\pi_3 (SU(3)/SU(2)) = 0$. The reduction of the $SU(3)$ link matrices
to $SU(2)$ link matrices is done by a coset decomposition \cite{GLSW}
$U(x,\mu) = \omega(x,\mu) \tilde{U}(x,\mu)$, where $U(x,\mu) \in SU(3)$,
$\tilde{U}(x,\mu) \in SU(2)$. In order to be able to use a geometric
definition of ${\rm Tr} F_{\mu \nu} \tilde{F}_{\mu \nu}$ on the $SU(2)$
matrices, we must make the latter as smooth as possible. This is done
by fixing to a maximal $SU(2)$ gauge, which amounts
to minimizing
$\sum_{x,\mu} (3-{\rm Re Tr}\, \omega (x,\mu))$ %
\footnote{Note that this
gauge is manifestly renormalizable.}.
For the minimizing procedure we
apply a combination of Metropolis and overrelaxation steps. We have
checked for most of our configurations that this results in a gauge
invariant ${\rm Tr} F_{\mu \nu} \tilde{F}_{\mu \nu}$.

\begin{figure}[tb]
\vspace{-0.75cm}
%\vspace{1.50cm}
%\begin{center}
\centerline{\psfig{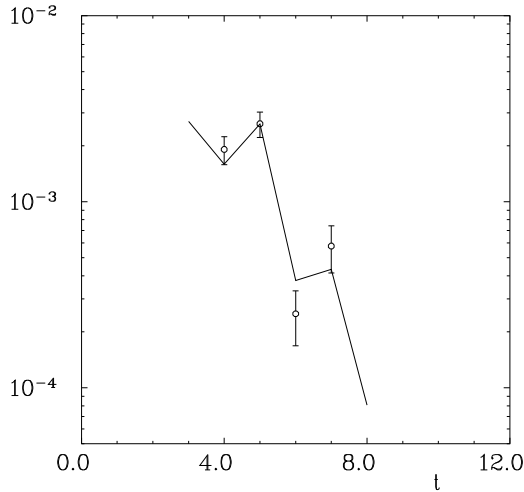}}
%\centerline{\psgraphic{amsfig}{73mm}{85mm}{0mm}{a}}
%\psgraphic{psfig}{8cm}{8cm}{0cm}{h}
%\vspace{-0.5cm}
\vspace{-1.25cm}
\caption{The correlation function (14), averaged over lattice momenta
${\vp} = \pm \, {\ve}_i \, 2\pi/16$ and spins ${\vs} = \pm \, {\ve}_i
\, |{\vs}|$, $i = 1, 2, 3$, where ${\ve}_i$ is the unit vector in
$i$-direction, as a function of $t$. Also shown is a two-parameter
fit of eq. (16) to the data, the fit interval being from $t = 4$ to
$t = 7$. For $t \geq 8$  the data is too noisy to be useful.}
%\end{center}
\label{figp}
\end{figure}

For $\bar{B}$ we take the wall source, while for $B$ we take the
ordinary local baryon operator (i.e. $\# 1$ in ref. \cite{MTC}). The
current is placed at $x_4 = 4$.
The result for the correlation function
(14), averaged over all possible momentum and spin combinations, is
shown in fig. 1. The data is fitted by the function (16) with
$A_{+}$, $A_{-}$, $m_N$ and $m_\Lambda$ taken from a fit \cite{MTC}
of $<B_{{\vz}}(t) \bar{B}_{{\vz}}(0)>$. The result is
indicated by the solid line. It leads to
\begin{equation}
\Delta \Sigma = 0.18 \pm 0.02,
\end{equation}
while the quark spin fraction of the $\Lambda$ comes out to be
$\Delta \Sigma_\Lambda = 0.22 \pm 0.04$.

In order to check our results for systematic errors, we have repeated
the calculation for $x_4 = 3$. We found the same result within errors.
Thus we can assume that $x_4 = 4$ is far enough away from the source,
so that the excited states have died out. Since our calculation involves
slowly moving (infrared) modes, it is important to test the efficiency
of the algorithm. We have found an autocorrelation time of
$\approx 25$ trajectories \cite{full}.
This translates into an autocorrelation time of five configurations.

\section{DISCUSSION}

Our result (17) lies within the errors of the experimental value
(3). One might object that our calculations are not
entirely realistic, as we work with four light quarks instead
of three and the nucleon matrix element is computed in the chiral
limit. But we have reasons to believe that none of these approximations
will have a significant
effect. As far as the flavor dependence is concerned,
one expects $\Delta \Sigma \propto \sqrt{N_f}$, which leads to a small
correction only.
The quark mass and momentum dependence, on the other hand,
is controlled by the $\eta^\prime$ mass which is large compared to
the pion mass.

\section*{ACKNOWLEDGEMENTS}

This work was supported in part by the Deutsche Forschungsgemeinschaft.
The numerical computations were performed on the Cray Y-MP at the
HLRZ. We thank both
institutions for their support. We would also like to thank M. L.
Laursen and U.-J. Wiese for their collaboration on various projects
of lattice topology, out of which some of the programs used in this work
developed. We are furthermore indebted to M. L\"uscher for pointing out
an error in the original manuscript.


\begin{thebibliography}{9}

\bibitem{EMC}
J. Ashman {\it et al.}, Phys. Lett. {\bf B206} (1988) 364; Nucl.
Phys. {\bf B328} (1990) 1

\bibitem{many}
An incomplete list contains: \\
S. J. Brodsky, J. Ellis and M. Karliner, Phys. Lett. {\bf B206} (1988)
309; \\
G. Altarelli and G. G. Ross, Phys. Lett. {\bf B212} (1988) 391; \\
A. V. Efremov and O. V. Teryaev, Dubna preprint E2-88-287 (1988); \\
R. D. Carlitz, J. C. Collins and A. H. Mueller, Phys. Lett. {\bf B214}
(1988) 229; \\
G. Veneziano, Mod Phys. Lett. {\bf A4} (1989) 1605; \\
R. L. Jaffe and A. Manohar, Nucl. Phys. {\bf B337} (1990) 509; \\
A. V. Efremov, J. Soffer and O. V. Teryaev, Nucl. Phys. {\bf B346}
(1990) 97; \\
H. Fritzsch, Phys. Lett. {\bf B229} (1989) 122; {\it ibid.} {\bf B242}
(1990) 451; \\
G. M. Shore and G. Veneziano, Phys. Lett. {\bf B244} (1990) 75 \\
See also: \\
K.-F. Liu, talk given at this conference

\bibitem{Ellis}
See also: \\
J. Ellis, M. Karliner and C. Sachrajda, Phys. Lett. {\bf B231} (1989)
497; \\
R. L. Jaffe and A. Manohar, ref. 2

\bibitem{Fritzsch}
See also: \\
H. Fritzsch, Max-Planck-Institut preprint MPI-Ph/91-34 (1991)

\bibitem{Mandula}
J. E. Mandula, Phys. Rev. Lett. {\bf 65} (1990) 1403; Nucl. Phys.
{\bf B} (Proc. Suppl.) {\bf 26} (1992) 365

\bibitem{MTC}
R. Altmeyer, K. D. Born, M. G\"ockeler,
R. Horsley, E. Laermann and G. Schierholz,
HLRZ preprint 92-17 (1992), to be published in Nucl. Phys. {\bf B}

\bibitem{Luscher}
M. L\"uscher, Commun. Math. Phys. {\bf 85} (1982) 39

\bibitem{P&S}
A. Phillips and D. Stone, Commun. Math. Phys. {\bf 103} (1986) 599;
{\it ibid.} {\bf 131} (1990) 255

\bibitem{Teper}
D. J. R. Pugh and M. Teper, Phys. Lett. {\bf B224} (1989) 159

\bibitem{GKLSW}
M. G\"ockeler, A. S. Kronfeld, M. L. Laursen, G. Schierholz and U.-J.
Wiese, Phys. Lett. {\bf B233} (1989) 192

\bibitem{full}
For a full account see: \\
R. Altmeyer, M. G\"ockeler, R. Horsley, E. Laermann and G. Schierholz,
in preparation

\bibitem{Fox}
I. A. Fox, J. P. Gilchrist. M. L. Laursen and G. Schierholz, Phys. Rev.
Lett. {\bf 54} (1985) 749

\bibitem{Choquet}
Y. Choquet-Bruhat, C. DeWitt-Morette and M. Dillard-Bleick, {\it
Analysis, Manifolds and Physics}, pp. 381-385, North-Holland (Amsterdam,
1982)

\bibitem{GLSW}
M. G\"ockeler, M. L. Laursen, G. Schierholz and U.-J. Wiese, Commun.
Math. Phys. {\bf 107} (1986) 467

\end{thebibliography}
\end{document}